\documentstyle[12pt]{article}
\begin{document}
\title{Theory of Weak Hypernuclear Decay$^*$}
\author{John F. Dubach,$^a$ Geoffrey B. Feldman,$^b$ Barry R. 
Holstein$^{a,c}$\\[5mm]
${}^a$Department of Physics and Astronomy\\
University of Massachusetts\\
Amherst, MA  01003\\
${}^b$Departamento de F\'{\i}sica\\
Colegio Universitario Tecnol\'{o}gico de Arecibo\\
Call Box 4010\\
Arecibo, Puerto Rico  00613\\
${}^c$Institute for Nuclear Theory\\
University of Washington\\
Seattle, WA  98195\\
and\\
Lorenzo de la Torre\\
Departamento di Fisica\\
Universidad di Antioquia\\
Medellin, Colombia}

\begin{titlepage}
\maketitle
\begin{abstract}
The weak nonmesonic decay of $\Lambda$-hypernuclei is studied in the context
of a one-meson-exchange model.  Predictions are made for the decay rate, the
p/n stimulation ratio and the asymmetry in polarized hypernuclear decay.
\end{abstract}
\vfill
$^*$Research supported in part by the National Science Foundation and the
Department of Energy.
\end{titlepage} 
\section{Introduction}

A great deal of attention has been focused over the last decade or so 
on the properties of $\Lambda$-hypernuclei, the study of which has yielded a
rich store of valuable information concerning the $\Lambda-N$ 
interaction.\cite{1}  Typically such hypernuclei are produced via the 
($K^-, \pi^-$) or ($\pi^+, K^+$) reaction, employing
kinematics wherein the resulting $\Lambda$ is produced with relatively 
low momentum.  The $\Lambda$ is generally
not formed in the hypernuclear ground state, but rather proceeds there 
via emission of a series of $\gamma$-rays, the study of which yields the 
hypernuclear levels.  However, the purpose of the present paper is not to 
study of this cascade process, but rather to analyze 
what happens once the $\Lambda$ has finally reached its ground 
($1s_{1\over 2}$) state.  In fact, the $\Lambda$ then
decays weakly, and there are intriguing aspects of this process which form the 
topic of this paper.

The properties of the lambda hyperon are familiar.  Having a mass
of 1116 MeV, zero isospin and unit negative strangeness, it decays nearly 100\% of the time
via the nonleptonic mode $\Lambda\rightarrow N\pi$ and details can be
found in the particle data tables\cite{2}
\begin{equation}
\Gamma_\Lambda={1\over 263\,\mbox{ps}}\qquad \mbox{B.R.}\,\,\Lambda\rightarrow
\left\{\begin{array}{cc}
p\pi^- & 64.1\% \\
n\pi^0 & 35.7\%
\end{array}\right.
\end{equation}
The decay can be completely described in terms of an effective Lagrangian with
two phenomenological parameters
\begin{equation}
{\cal H}_w=g_w\bar{N}(1+\kappa\gamma_5)\vec{\tau}\cdot\vec{\pi}\Lambda
\end{equation}
where $g_w=2.35\times 10^{-7},\kappa=-6.7$ and $\Lambda$ 
is defined to occupy the
lower entry of a two component column isospinor 
s---$\Lambda\equiv \Lambda s$ with $s=\left(\begin{array}{c} 0\\ 1
\end{array}\right).$

However, it was realized early on by Primakoff and Cheston that when the 
$\Lambda$ is bound in a hypernucleus, its decay properties are altered
dramatically.\cite{3}  The kinematics for free $\Lambda\rightarrow
N\pi$ decay at rest give the energy and momentum of the final state nucleon
\begin{equation}
T_N={(m_\Lambda-m_N)^2-m_\pi^2\over 2m_\Lambda}\approx 5\,\,{\rm MeV},\quad 
p_N=\sqrt{(T_N+m_N)^2-m_N^2}\approx 100 \,\,{\rm MeV/c}
\end{equation}
Thus, $p_N$ is less than the nuclear Fermi momentum for all but the lightest
nuclei, and the $N\pi$ decay of a $\Lambda$-hypernucleus is thus Pauli blocked.
(Actually 
this suppression is even stronger than indicated above since
typically the $\Lambda$ is bound by 5-25 MeV.)

A simple estimate of this suppression is given within a simple shell model, 
wherein, taking free space kinematics (neglecting any effects of binding 
energy or of wavefunction 
distortion), and recognizing the fact that the pion recoils against the
nucleus as a whole instead of a single nucleon, one finds the simple
expression\cite{50}
\begin{equation}
{1\over \Gamma_\Lambda}\Gamma_{\Lambda\rightarrow N\pi}=
1-{1\over 2}\sum_{nj\ell}N_{nj\ell} |\langle
nj\ell|j_\ell(k_\pi r)|1s_{1\over 2}\rangle|^2
\end{equation}
where $N_{nj\ell}$ is the occupation number for the indicated state and 
$k_\pi\sim$ 100 MeV/c is the pion momentum.  The
result of this calculation reveals that the importance of such pionic decays
rapidly falls as a function of nuclear mass, as shown in Figure 1.\cite{34}
\begin{figure}
\vspace{3.5in}

\caption{Ratio of calculated rate of $N\pi$ decay of 
hypernuclei to that of a free
$\Lambda$ as a function of nuclear mass number.}
\end{figure}

However, while the existence of the nuclear medium suppresses the 
$N\pi$ mode, it also opens
up a completely new possibility, that of the nucleon-stimulated 
decay---$\Lambda N\rightarrow NN$---which is, of course, unavailable to a 
free $\Lambda$.\cite{4}  This reaction is the $\Delta S=1$ analog of the weak
$NN\rightarrow NN$ reaction responsible for nuclear parity violation,\cite{5} 
but with the difference that the weak parity-conserving 
$\Lambda N\rightarrow NN$ decay is {\it also} observable (in the 
$NN\rightarrow NN$ case the weak parity-conserving component is, of course,
dwarfed by the strong interactions.)  Note that the energy and momentum
available in this process are, if shared equally by both outgoing nucleons,
\begin{equation}
T_N\approx{1\over 2}(m_\Lambda -m_N)\approx 90\,\,{\rm MeV},\quad
p_N=\sqrt{(T_N+m_N)^2-m_N^2}\approx 420\,\,{\rm  MeV/c}
\end{equation}
which is well above a typical Fermi energies and momenta.  Thus the Pauli effect 
does not significantly suppress the nonmesonic mode, and consequently the 
importance of the nonmesonic (NM) channel compared to its mesonic counterpart
is expected to increase rapidly with A.  This prediction is fully  borne
out experimentally, as shown in Figure 2.  It is evident that once $A\geq 10$
the mesonic decay becomes but a small fraction of the dominant nonmesonic 
process.\cite{34}
\begin{figure}
\vspace{3.5in}
\caption{Measured ratio of nonmesonic ($\Lambda N\rightarrow NN$) to mesonic
($\Lambda\rightarrow N\pi$) hypernuclear decay as a function of nuclear mass
number.}
\end{figure}

The dominant mode of hypernuclear weak decay then is not the 
pionic mode favored by a free $\Lambda$ but becomes rather the far more
complex $\Lambda N\rightarrow NN$ process.\cite{6}  The observables which can be
measured experimentally and should be confronted with theoretical predictions 
include
\begin{itemize}
\item[i)] the overall nonmesonic decay rate $\Gamma_{NM}$;
\item[ii)] the ratio of proton-stimulated ($\Lambda p\rightarrow np$) to
neutron-stimulated ($\Lambda n\rightarrow nn$) 
decay---$\Gamma^p_{NM}/\Gamma^n_{NM}
\equiv \Gamma_{NM}(p/n)$;
\item[iii)] the ratio of parity-violating to parity-conserving
decay---$\Gamma^{PV}_{NM}/\Gamma^{PC}_{NM}\equiv\Gamma_{NM}(PV/PC)$---which 
is measured, {\it e.g.},
via the proton asymmetry in polarized hypernuclear decay;
\item[iv)] final state n,p decay spectra;
\item[v)] {\it etc.}
\end{itemize}

The present experimental situation is somewhat limited. Most of the early
experiments in the field employed bubble chamber or emulsion techniques.
It was therefore relatively straightforward to determine the ratio of the
decay rates of the two modes, but much more difficult to measure the
absolute rates.  This changed when an early Berkeley 
measurement on ${}^{16}_\Lambda \mbox{O}$ yielded the value\cite{7}
\begin{equation}
{\Gamma({}^{16}_\Lambda \mbox{O})\over \Gamma_\Lambda}=3\pm 1
\end{equation}
However, this was still a very low statistics experiment with sizable 
background contamination.  Recently a CMU-BNL-UNM-Houston-Texas-Vassar 
collaboration undertook a 
series of  direct timing---fast counting---hypernuclear lifetime measurements 
yielding the results summarized in Table 1.\cite{8}

\begin{table}
\begin{center}
\begin{tabular}{|c|c|c|c|}\hline
  &${}^5_\Lambda \mbox{He}$ &$ {}^{11}_\Lambda \mbox{B}$ & ${}^{12}_\Lambda 
\mbox{C} $\\ 
\hline
\hline
${1\over \Gamma_\Lambda}\Gamma_{NM}$ & $0.41\pm 0.14$& $ 1.25\pm 0.16^*$ & 
$1.14\pm 0.20$ \\ \hline
$\Gamma_{NM}(p/n)$ &$1.07\pm 0.58 $  &$0.96^{+0.8}_{-0.4}$ &$0.75^{+1.5}_{-0.35}$ \\ \hline
\end{tabular}
\end{center}
\caption{Experimental BNL data for nonmesonic hyperon decay.  $^*$Note that we
have scaled the experimental number in order to remove the mesonic decay 
component.}
\end{table}
In addition, there exist a number of older emulsion measurements in light
($A\leq 5$) hypernuclei, details of which can be found in a recent review 
article.\cite{6}  The only experimental numbers for heavy 
systems are obtained from delayed fission measurements on hypernuclei
produced in $\bar{p}$-nucleus collisions and are of limited statistical
precision\cite{9}
\begin{equation}
\tau({}^{238}_{\Lambda}\mbox{U})=(1^{+1}_{0.5})\times 10^{-10}\,\,\mbox{sec}.\qquad
\tau({}^{209}_{\Lambda}\mbox{Bi})=(2.5^{+2.5}_{-1.0})\times 10^{-10}\,\,\mbox{sec}.
\end{equation}

The theoretical problem of dealing
with a weak two-body interaction within the nucleus has been faced previously 
in the
context of nuclear parity violation, and one can build on what has been learned
therein.\cite{10}  Specifically, the weak interaction at the quark level is 
shortranged, involving W,Z-exchange.  However, because of the hard core 
repulsion the
NN effects are modelled in terms of long-range one-meson
exchange interaction, just as in the case of the conventional strong 
nucleon-nucleon interaction,\cite{11} but now 
with one vertex being weak and parity-violating while
the second is strong and parity-conserving---{\it cf.} Figure 3.  
The exchanged mesons are 
the lightest ones---$\pi^\pm ,\rho ,\omega$---associated with the longest
range. (Exchange of 
neutral spinless mesons is forbidden by Barton's theorem.\cite{12})
\begin{figure}
\vspace{2.5in}
\caption{Meson exchange picture of nuclear parity violation.}
\end{figure}

A similar picture of hypernuclear decay can then be constructed, but with
important differences.  While the basic meson-exchange diagrams appear as
before, the weak vertices must now include both parity-conserving {\it and}
parity-violating components, and the list of exchanged mesons must be expanded 
to include
both neutral spinless objects ($\pi^0,\eta^0$) as well as strange mesons
($K,K^*$), as first pointed out by Adams.\cite{13}  Thus the problem is 
considerably more 
challenging than the corresponding and already difficult issue of nuclear 
parity violation.

One of the significant problems in such a calculation involves the evaluation
of the various
weak amplitudes.  Indeed, the only weak couplings which are completely
model-independent are those involving pion emission, which are given in Eqn. 2.
In view of this, a number of calculations have included {\it only} this 
longest range component.
Even in this simplified case, however, 
there is considerable model-dependence, as the results are
strongly sensitive to the short-ranged correlation function assumed for the
nucleon-nucleon interaction, as will be seen.  
Below we shall review previous theoretical work in this area and detail our
own program, which involves a systematic quark model- (symmetry-) based
evaluation of weak mesonic couplings to be used in hypernuclear decay 
calculations.

A brief outline of our paper is as follows.  In section 2 we employ the quark
model in order to construct the weak potentials which will be used to study
the nonmesonic decay process.  In sections 3,4 we apply these potentials in the
regime of nuclear matter, finite nuclei respectively.  Finally, we summarize 
our results in a concluding section 5.

\section{Hypernuclear Decay: Effective Interaction}

As discussed above, a primary challenge in the theoretical analysis of 
hypernuclear weak decay is in the calculation of the weak baryon-baryon-meson
(BB'M) vertices, since the strong couplings are known reasonably well either
from direct measurement or from the approximate validity of vector dominance.  
However, in the case of the weak interaction only $\Lambda N\pi$ vertices 
are accessible experimentally---evaluation of any remaining weak couplings
requires a model.  Our approach is based on the quark model together with a 
generalization of techniques previously developed to deal with nonleptonic 
processes.\cite{14}  

We begin with the parity-conserving weak interaction, which we evaluate
using a pole model and the diagrams shown in Figure 4.
What is required then is the evaluation of two-body matrix elements
\begin{equation}
\langle B'|{\cal H}_w^{(+)}|B\rangle ,\langle P'|{\cal H}_w^{(+)}|P\rangle ,
\langle V'|{\cal H}_w^{(+)}|V\rangle
\end{equation}
where ${\cal H}_w^{(+)}$ is the parity-conserving $\Delta S=1$ weak 
Hamiltonian and B,P,V represent baryons, pseudoscalar, vector mesons 
respectively.  Although quark model based, the clearest way to characterize 
our results is in terms of the symmetry SU(6)$_w$ which is a generalization 
of SU(6) allowing a relativistically correct theory for arbitrary boosts 
along one direction $\hat{z}$.\cite{15}  This is sufficient for a full 
treatment of weak nonleptonic vertices.  The first applications to weak 
phenomenology were those of McKellar and Pick\cite{16} and of Balachandran 
et al.,\cite{17}  while a comprehensive SU(6)$_w$ treatment of nuclear 
parity violation was presented in ref.10.  
\begin{figure}
\vspace{3in}
\caption{Pole model diagrams used 
in evaluation of the parity conserving component of
$\Lambda N\rightarrow NN$.}
\end{figure}

We begin by writing in SU(3) 
tensor notation the $\Delta S=1$ weak nonleptonic Hamiltonian
\begin{equation}
{\cal H}_w(\Delta S=1)={G_v\over 2\sqrt{2}}\cos\theta_c\sin\theta_c
\{J_{\mu 1}^2,J^{\mu 1}_3\}_+ +{\rm h.c.}
\end{equation}
where $J_{\mu j}^i=(V+A)_{\mu j}^i$ is the weak hadronic current with SU(3) 
indices i,j, $G_v\approx 10^{-5}
m_p^{-2}$ is the conventional weak coupling constant, and $\theta_c$ is 
the Cabibbo angle.  Now following the procedure of Balachandran et
al.\cite{17} we express the vector and axial currents in terms of SU(6)$_w$ 
currents P,Q,R,S
\begin{equation}
\begin{array}{ll}
A_{+b}^a=-R^{2a}_{2b-1} & A^a_{-b}=-R^{2a-1}_{2b} \nonumber\\
V^a_{+b}=iQ^{2a}_{2b-1} & V^a_{-b}=-iQ^{2a-1}_{2b}\nonumber\\
V^a_{3b}={1\over 2}(P^{2a}_{2b}+P^{2a-1}_{2b-1})  & A^a_{0b}={1\over 2}
(P^{2a}_{2b}-P^{2a-1}_{2b-1})\nonumber\\
V^a_{0b}={1\over 2}(S^{2a}_{2b}+S^{2a-1}_{2b-1}) &  A^a_{3b}={1\over 2}
(S^{2a}_{2b}-S^{2a-1}_{2b-1})
\end{array}
\end{equation}
The A,V indices a,b are SU(3) indices while on P,Q,R,S they represent 
SU(6)$_w$ indices.  For the weak parity-conserving Hamiltonian we find
\begin{equation}
{\cal H}_w^{(+)}(\Delta S=1)={G_v\over \sqrt{2}}\cos\theta_c\sin\theta_c
{}^{(+)}{\cal O}^{56}_{43}
\end{equation}
where
\begin{eqnarray}
{}^{(+)}{\cal O}^{AB}_{CD}&=&U^{\{1,B\}}_{\{D,2\}}+U^{\{2,A\}}_{\{C,1\}}
+U^{[1,B]}_{[D,2]}+U^{[2,A]}_{[C,1]}\nonumber\\
&+&W^{\{2,A\}}_{\{D,2\}}+W^{\{1,B\}}_{\{C,1\}}+W^{[2,A]}_{[D,2]}+W^{[1,B]}_{[C,1]}
+W^A_D+W^B_C
\end{eqnarray}
with
\begin{eqnarray}
U^{AB}_{CD}&=&S^A_CS^B_D-P^A_CP^B_D\nonumber\\
W^{AB}_{CD}&=&-R^A_CR^B_D-Q^A_CQ^B_D
\end{eqnarray}
and $[ , ]$ and $\{ , \}$ represent respective 
antisymmetrization and symmetrization of 
SU(6)$_w$ indices.  Thus ${\cal H}_w^{(+)}(\Delta S=1)$ transforms as 
$\{405\} + \{189\} + \{35\}$.  We can then identify 
the various ways in which to 
couple this Hamiltonian to two baryons or mesons.
\begin{eqnarray}
\alpha: \qquad && [M_{35} \times M_{35}]_{35}\nonumber\\
\beta:\qquad  && [M_{35} \times M_{35}]_{189}\nonumber\\
\gamma:\qquad &&  [M_{35} \times M_{35}]_{405}\nonumber\\
\delta:\qquad &&  [\bar{B}_{56} \times B_{56}]_{35}\nonumber\\
\epsilon:\qquad &&  [\bar{B}_{56} \times B_{56}]_{405}
\end{eqnarray}
Specifically, we find
\begin{eqnarray}
\langle K^0|{\cal H}_w^{(+)}|\pi^0\rangle &=&-{1\over \sqrt{2}}\alpha\nonumber\\
\langle K^+|{\cal H}_w^{(+)}|\pi^+\rangle &=&\alpha -\beta\nonumber\\
\langle K^0|{\cal H}_w^{(+)}|\eta^0\rangle &=&-{1\over \sqrt{6}}\alpha\nonumber\\
\langle K^{*0}(0)|{\cal H}_w^{(+)}|\omega^0(0)\rangle
&=&\langle K^{*0}(\uparrow)|{\cal H}_w^{(+)}|\omega^0(\uparrow)\rangle=
-{1\over \sqrt{6}}\alpha\nonumber\\
\langle K^{*0}(0)|{\cal H}_w^{(+)}|\rho^0(0)\rangle &=&
\langle K^{*0}(\uparrow)|{\cal H}_w^{(+)}|\rho^0(\uparrow)\rangle
=-{1\over \sqrt{2}}\alpha\nonumber\\
\langle K^{*+}(0)|{\cal H}_w^{(+)}|\rho^+(0)\rangle&=&\alpha+\beta\nonumber\\
\langle K^{*+}(\uparrow)|{\cal H}_w^{(+)}|\rho^+(\uparrow)\rangle&=&\alpha+\gamma
\end{eqnarray}

In order to proceed we shall assume the validity of the $\Delta I={1\over 2}$ 
rule for the $\Delta S=1$ weak Hamiltonian.  There has, of course, been a 
great deal of theoretical work attempting to identify the origin of this 
result.\cite{18}  The best indication at present is that in the case of 
hyperon decay, the suppression of $\Delta I={3\over 2}$ effects is associated 
with the so-called Pati-Woo theorem,\cite{19} which guarantees the vanishing of
$\langle B'|{\cal H}_w^{(+)}|B\rangle$ between color-singlet baryons, while 
in the case of kaon transitions the validity of the $\Delta I={1\over 2}$ rule 
appears to be associated with presently incalculable long distance 
contributions.\cite{20}  It is not our purpose here to enter into the debate 
concerning the dynamical origin 
of the $\Delta I={1\over 2}$ rule, only to note its
existence in other $\Delta S=1$ nonleptonic weak processes and to {\it assume} 
its validity in the realm of nonmesonic hypernuclear decay 
as one of our inputs. (We shall return
to this issue later.)  This may be done through the introduction of the 
iso-spurion $s=\left(\begin{array}{c}
0\\1
\end{array}\right)$.  Then we can write
\begin{equation}
\langle K|{\cal H}_w^{(+)}|\pi\rangle =A_{K\pi}K^\dagger\vec{\tau}\cdot
\vec{\pi} s
\end{equation}
so that
\begin{equation}
\langle K^+|{\cal H}_w^{(+)}|\pi^+\rangle =\sqrt{2}A_{K\pi}=-\sqrt{2}
\langle K^0|{\cal H}_w^{(+)}|\pi^0\rangle
\end{equation}
Comparison with eqn. 15 yields
\begin{equation}
\alpha=\sqrt{2}A_{K\pi}\qquad\beta=0
\end{equation}
Similar considerations for the vector meson matrix element $\langle
K^*|{\cal H}_w^{(+)}|\rho\rangle$ give the additional result
\begin{equation}
\gamma=0
\end{equation}
Thus, all two-body weak mesonic amplitudes are determined in terms of 
the single parameter $A_{K\pi}$, which can in turn be related via PCAC to 
the experimental amplitude for $K\rightarrow \pi\pi$ decay.\cite{21}  The 
only subtlety is that it is essential to take into account the momentum 
dependence of the three-body matrix elements required by simultaneous 
consistency with current algebra limits and with the symmetry requirement
that $\langle \pi\pi|{\cal H}_w^{(-)}|K\rangle$ must vanish in the SU(3) 
limit.\cite{22}  For the decay $K^0\rightarrow\pi^0\pi^0$ this implies
\begin{equation}
A(K^0\rightarrow\pi^0\pi^0\rangle\propto 2k^2-q_{01}^2-q_{02}^2
\end{equation}
Then we find
\begin{eqnarray}
\lim_{q_{02}\rightarrow 0}\langle \pi^0_{q_{01}}\pi^0_{q_{02}}|{\cal H}_w^{(-)}
|K^0_k\rangle ={-i\over F_\pi}\langle \pi^0_{q_{01}}|[F^5_{\pi^0},
{\cal H}_w^{(-)}]|K^0_k\rangle\nonumber\\
=-{i\over 2F_\pi}\langle \pi^0_{q_{01}}|{\cal H}_w^{(+)}|K^0_k\rangle
={k\cdot q_{01}\over 2(m_K^2-m_\pi^2)}\langle \pi^0\pi^0|{\cal H}_w^{(-)}
|K^0\rangle_{\rm expt.}
\end{eqnarray}
so that
\begin{equation}
A_{K\pi}\simeq -iF_\pi{k\cdot q_{0}\over m_K^2-m_\pi^2}\langle\pi^0\pi^0|
{\cal H}_w^{(-)}|K^0\rangle_{\rm expt.}\equiv\tilde{A}_{K\pi}k\cdot q_0
\end{equation}

Having determined the two-body parity-conserving meson amplitudes,
we turn to the corresponding baryon matrix elements.  Here the Pati-Woo 
theorem guarantees the validity of the $\Delta I={1\over 2}$ rule, and 
we can characterize the matrix elements in terms of two independent parameters 
$A_{N\Lambda},A_{N\Sigma}$ defined via
\begin{equation}
\langle N|{\cal H}_w^{(+)}|\Lambda\rangle =A_{N\Lambda}\bar{N}\Lambda,\qquad
\langle N|{\cal H}_w^{(+)}|\Sigma\rangle=A_{N\Sigma}\bar{N}\vec{\tau}
\cdot\vec{\Sigma} s
\end{equation}
These quantities can be determined via current algebra/PCAC as before
\begin{eqnarray}
\lim_{q_0\rightarrow 0}\langle \pi^0n|{\cal H}_w^{(-)}|\Lambda\rangle &=&
{-i\over F_\pi}\langle n|[F^5_{\pi^0},{\cal H}_w^{(-)}]|\Lambda\rangle =
{i\over 2F_\pi}\langle n|{\cal H}_w^{(+)}|\Lambda\rangle\nonumber\\
\lim_{q_0\rightarrow 0}\langle \pi^0p|{\cal H}_w^{(-)}|\Sigma^+\rangle&=&
{-i\over F_\pi}\langle p|[F^5_{\pi^0},{\cal H}_w^{(-)}]|\Sigma^+\rangle=
{i\over 2F_\pi}\langle p|{\cal H}_w^{(+)}|\Sigma^+\rangle
\end{eqnarray}
Then assuming no momentum dependence for the baryon S-wave decay 
amplitude\cite{2}
\begin{eqnarray}
A_{N\Lambda}\simeq -iF_\pi\langle\pi^0n|{\cal H}_w^{(-)}|\Lambda\rangle
=-i4.46\times 10^{-5}\,\,\mbox{MeV}\nonumber\\
A_{N\Sigma}\simeq -i\sqrt{2}F_\pi\langle \pi^0p|{\cal H}_w^{(-)}|\Sigma^+\rangle
=i4.36\times 10^{-5}\,\,\mbox{MeV}.
\end{eqnarray}

Now although the specific technique used in order to obtain eqn. 15 was the 
symmetry SU(6)$_w$, it is straightforward to demonstrate that identical 
results are obtained in the simple valence quark model.  Indeed since the 
SU(6)$_w$ indices $A=1,2\ldots 6$ correspond to quark flavor spin 
states---$1=u\uparrow, 2=u\downarrow, \ldots 6=s\downarrow$---then the 
SU(6) 405 tensors $U^{AB}_{CD}$ contain four external indices which can be 
expressed by their action on four quark fields.  In particular the 
combination of tensors which appears in the Hamiltonian is the same as the 
nonrelativistic reduction of
\begin{equation}
H_U=V_0V_0+A_0A_0-V_3V_3-A_3A_3
\end{equation}
The W portion follows likewise from the transverse currents.  In forming 
tensors from the baryon and meson fields with the same transformation 
properties as a given tensor $U^{AB}_{CD}$ we must contract the SU(6)$_w$ 
indices that are not set equal to A, B, C, D.  Thus the meson 
to meson amplitudes require no contraction, but the baryon to baryon 
vertices involve a summation
\begin{equation}
[\bar{B}_{56}\times B_{56}]_{405}\sim\bar{B}^{eab}B_{ecd}
\end{equation}
Of course, several terms of this form must be symmetrized in order to
form the required Hamiltonian.  These observations can be combined to 
give a pictorial way in which to represent the group theoretical 
structure---instead of using SU(6)$_w$ we can utilize quark indices 
throughout.  Then the parameters a,b,...e correspond to the quark 
spin sums depicted in Figure 5.  
\begin{figure}
\vspace{2in}
\caption{Baryon-baryon matrix element of the weak Hamiltonian.}
\end{figure}

Baryons and mesons are described by SU(6) wavefunctions (which are equivalent 
to the tensors $B_{abc}$ and $M_{ab}$) and the wavy line describes 
the SU(6)$_w$ Hamiltonian.  A study of indices in Eqn. 14 and
Figure 4 reveals that the two approaches are equivalent.  At this stage 
then the diagrams could be considered merely as simple ways by which to 
obtain the SU(6)$_w$ transformation properties.  However, in a quark model 
they also have a well-defined dynamical meaning and can, in some 
approximation, be calculated.  It is not our purpose here to undertake 
such a calculation, however.  Indeed naive quark-only diagrams, which omit the
all-important gluon and/or quark sear content cannot possibly convey the
intricate dynamical details which characterize processes involving 
low energy strong interactions.  An archetypical example is the empirical
validity of the $\Delta I={1\over 2}$ rule in $\Delta S=1$ nonleptonic
processes such as $K\rightarrow 2\pi , 3\pi$ and $B\rightarrow B'\pi$.
Simple quark diagrams would suggest that the strength of the $\Delta I={1\over
2}$ and $\Delta I={3\over 2}$ components of the weak interaction should be
comparable.  Inclusion of leading-log gluonic corrections via renormalization
group methods provides an enhancement of the $\Delta I={1\over 2}$ piece over
its $\Delta I={3\over 2}$ counterpart by a factor of 3-4, and the remaining
factor of 5-6 required in order to agree with experiment arises (presumably)
from subtle details of the of the strongly interacting particles involved.
Calculations of such detail are well beyond the ability of present theoretical
methods and
consequently we shall simply utilize the semi-empirical 
results quoted in Eqns. 22, 25, which encode the details of strong and weak
interaction dynamics in a set of experimental constants.

We can now, in terms of these results, construct any of the needed 
amplitudes---$T_w^{\Lambda N\pi},
T_w^{NNK},T_w^{\Lambda N\rho},$ {\it etc}.---via the pole diagrams 
indicated in Figure 3.  As a check of the validity of 
this technique, we evaluate the only amplitude which is experimentally 
accessible---the P-wave $\Lambda N\pi$ decay amplitude.  We predict
\begin{eqnarray}
T_w^{\Lambda N\pi}&=&\bar{N}\vec{\tau}\cdot\vec{\pi} \Lambda\left[Q_{\Lambda\Sigma\pi}
{1\over m_N-m_\Sigma}A_{N\Sigma}\right.\nonumber\\
&&\left.+Q_{NN\pi}{1\over m_\Lambda-m_N}A_{N\Lambda}+Q_{\Lambda NK}{1\over 
m_K^2-m_\pi^2}\tilde{A}_{K\pi}m_\pi^2\right]
\end{eqnarray}
where the strong couplings $Q_{BB'M}$ are defined via use of the 
generalized Goldberger-Treiman relation and PCAC/SU(3).\cite{23}
\begin{equation}
Q_{B^iB^jM^a}={m_{B^i}+m_{B^j}\over \sqrt{2}F_\pi}[dd_{ija}+iff_{ija}]
\end{equation}
and\cite{24}
\begin{equation}
f+d=1.26,\qquad d/(f+d)\simeq 0.63
\end{equation}
from fits to semileptonic hyperon decay.  Using
\begin{eqnarray}
Q_{\Lambda\Sigma\pi}&=&{1\over \sqrt{3}}d{m_\Lambda+m_N\over F_\pi}\nonumber\\
Q_{NN\pi}&=&-(f+d){m_N\over F_\pi}\nonumber\\
Q_{\Lambda NK}&=&-{1\over 2\sqrt{3}}(3f+d){m_\Lambda+m_N\over F_\pi}
\end{eqnarray}
we find
\begin{equation}
T_w^{\Lambda n\pi^0}(\mbox{theo.})=1.35\times 10^{-6}
\end{equation}
to be compared with the experimental value\cite{2}
\begin{equation}
T_w^{\Lambda n\pi^0}(\mbox{expt.})=1.61\times 10^{-6}
\end{equation}

Actually this near agreement should be considered somewhat fortuitous.
Indeed it is well-known that use of $\langle B'|{\cal H}_w^{(+)}|
B\rangle$ matrix elements as determined from current algebra/PCAC and 
an SU(3) fit to S-wave hyperon decay amplitudes---$F=-2.4D=-0.92\times 
10^{-7}$---yields generally a rather poor fit to corresponding P-wave 
hyperon decay amplitudes, with values in some cases about 50\% too 
small.\cite{25}  However, a slight shift in these 
parameters---$F=-1.8D=-1.44\times 10^{-7}$---yields a much better fit and 
reproduces nearly all amplitudes to within 15\%.\cite{14} Since for our
purposes we require only the vertices $A_{N\Lambda},A_{N\Sigma}$
and since these differ only by about 15\% in the two parameterizations quoted 
above, we shall not worry about this difference, given the
preliminary nature of our investigation.  Indeed we have verified that 
calculations performed with either set of couplings are quite similar.

Having thus obtained a handle on the parity-conserving, or P-wave amplitudes, 
we move on to consider the parity-violating meson vertices, which are to
be inserted into the diagrams shown in Figure 6 in order to generate 
two-body $\Lambda N\rightarrow NN$ operators.
\begin{figure}
\vspace{2.5in}
\caption{Two body operators for parity-violating nonmesonic hypernuclear
decay.}
\end{figure}

Phenomenologically, assuming again only the $\Delta I={1\over 2}$ rule, 
we can write ${\cal H}_w^{(-)}$ in terms of eight unknown couplings
\begin{eqnarray}
{\cal H}_w^{(-)}&=&A_{\pi\Lambda N}\bar{N}\vec{\tau}\cdot
\vec{\pi} \Lambda+A_{\eta\Lambda N}
\bar{N} \eta\Lambda\nonumber\\
&+&C_{KNN}\bar{N}sK^\dagger N+D_{KNN}\bar{N}NK^\dagger s\nonumber\\
&+&A_{\rho\Lambda N}\bar{N}\gamma_\mu\gamma_5\vec{\tau}\cdot
\vec{\rho}^\mu \Lambda
+A_{\omega\Lambda N}\bar{N}\gamma_\mu\gamma_5\omega^\mu \Lambda\nonumber\\
&+&C_{K^*NN}\bar{N}s\gamma_\mu\gamma_5K^{*\mu\dagger}N
+D_{K^*NN}\bar{N}\gamma_\mu\gamma_5NK^{*\mu\dagger}s
\end{eqnarray}
These various couplings may be interrelated via the SU(6)$_w$ symmetry 
scheme, or equivalently by use of the quark model.  We find, in SU(6)$_w$, 
the Hamiltonian
\begin{equation}
{\cal H}_w^{(-)}(\Delta S=1)={G_v\over \sqrt{2}}\cos\theta_c\sin\theta_c
{}^{(-)}{\cal O}^{56}_{43}
\end{equation}
where
\begin{eqnarray}
{}^{(-)}{\cal O}^{AB}_{CD}&=&T^{[2,A]}_{\{C,1\}}+T^{\{2,A\}}_{[C,1]}
-T^{[1,B]}_{\{D,2\}}-T^{\{1,B\}}_{[D,2]}\nonumber\\
&+&V^{[2,A]}_{\{D,2\}}+V^{\{2,A\}}_{[D,2]}-V^{[1,B]}_{\{C,1\}}
-V^{\{1,B\}}_{[C,1]}+V^B_C-V^A_D
\end{eqnarray}
with
\begin{eqnarray}
T^{AB}_{CD}&=&P^A_CS^B_D-S^A_CP^B_D\nonumber\\
V^{AB}_{CD}&=&-iR^A_CQ^B_D+iQ^A_CR^B_D
\end{eqnarray}
Note that T describes the effects of currents along the boost direction 
(0,3), while V includes currents orthogonal to this direction 
(1,2 or +,--).  The weak Hamiltonian is seen to involve the representations
\begin{equation}
{\cal H}_w^{(-)}=280+\overline{280}+35.
\end{equation}
We can then define, in an obvious notation, the various ways in which
to couple the two baryons and meson together in a CP-invariant fashion
\begin{eqnarray}
a_T,a_V:\qquad && [(\bar{B}B)_{35}\times M_{35}]_{280,\overline{280}}\nonumber\\
b_T,b_V:\qquad  && [(\bar{B}B)_{405}\times M_{35}]_{280,\overline{280}}\nonumber\\
c_V:\qquad &&  [(\bar{B}B)_{35}\times M_{35}]_{35}
\end{eqnarray}
In terms of these couplings we have (temporarily, for simplicity, 
omitting $a_V,a_T$ contributions)
\begin{eqnarray}
A^{b,c}_{\pi\Lambda N}&=&-{1\over \sqrt{6}}({1\over 6}b_V-{1\over 12} b_T
-{1\over 2}c_V)\nonumber\\
A^{b,c}_{\rho\Lambda N}&=&{1\over \sqrt{6}}(-{1\over 6}b_V+{1\over 4}b_T
+{1\over 2}c_V)\nonumber\\
A^{b,c}_{\omega\Lambda N}&=&{1\over 36\sqrt{2}}(4b_V-5b_T-6c_V)\nonumber\\
C^{b,c}_{KNN}&=&-{1\over \sqrt{2}}({1\over 36}b_V+{1\over 36}b_T-{1\over 3}c_V)
\nonumber\\
D^{b,c}_{KNN}&=&\sqrt{2}(-{5\over 36}b_V+{1\over 9}b_T+{1\over 6}c_V)
\nonumber\\
C^{b,c}_{K^*NN}&=&{1\over 9\sqrt{2}}(2b_V-b_T-5c_V)\nonumber\\
D^{b,c}_{K^*NN}&=&{1\over 9\sqrt{2}}(-{1\over 2}b_V+b_T+c_V)\nonumber\\
A^{b,c}_{\eta\Lambda N}&=&\sqrt{3}A_{\pi\Lambda N}
\end{eqnarray}
where for consistency with Lorentz invariance it is necessary to require 
$b_V=-b_T$.\cite{10}  Then $b_V(=-b_T)$ and $c_V$ are determined from a 
fit to parity-violating hyperon decay
\begin{equation}
b_V=-b_T=-1.28c_V=-4.0\times 10^{-7}
\end{equation}

We now return to the couplings $a_V,a_T$.  It has been shown in ref. 16 
that such $a$-type coupling constants correspond to the so-called 
``factorization" terms\cite{26} wherein the meson is coupled to the 
vacuum by one of the weak currents in all possible ways consistent 
with Fierz reordering.\cite{27}  Such a term is very small in the case of 
pion emission because of the near conservation of the axial current.  
However, the factorization term is substantial in the case of 
vector meson vertices and, in fact, originally was the only term considered 
to contribute to vector meson emission.  A naive evaluation of such 
terms explicitly violates the $\Delta I={1\over 2}$ rule, as 
factorization reflects only the basic symmetry of the Hamiltonian.  These 
couplings have been determined in ref. 16 as
\begin{equation}
a_T={1\over 3}a_V={3\over 5}G_v\cos\theta_c\sin\theta_c\langle\rho|V^3_\mu
|0\rangle\langle p|A^{3\mu}|p\rangle .
\end{equation}
Since $\Delta I={3\over 2}$ effects are suppressed with respect to their 
$\Delta I={1\over 2}$ counterparts by a factor of twenty or so, 
we shall neglect them in this preliminary analysis of hypernuclear
decay.\footnote{As argued above, this suppression does not follow in the
simple quark model but rather requires consistent treatment of soft gluon
effects.  Thus, for example, recent ``direct" quark calculations\cite{44,45} 
predict a substantial $\Delta I={3\over 2}$ components but are open to
question since they omit these important strong interaction effects.  Thus in
our approach we instead employ the empirical results and explicitly omit
$\Delta I={3\over 2}$ terms.}  
We then find the factorization contributions to the various couplings
\begin{eqnarray}
A^a_{\pi\Lambda N}&=&C^a_{KNN}=D^a_{KNN}=A^a_{\eta\Lambda N}=0\nonumber\\
A^a_{\rho\Lambda N}&=&\sqrt{3}a_T\nonumber\\
A^a_{\omega\Lambda N}&=&-{1\over 3}a_T\nonumber\\
C^a_{K^*NN}&=&{30\over 9}a_T\nonumber\\
D^a_{K^*NN}&=&{8\over 9}a_T.
\end{eqnarray}

Having determined values for each of the weak couplings we can insert 
them into the meson exchange diagrams indicated in Figure 6 in order to 
determine the various weak $\Lambda N\rightarrow NN$ ``potential" forms.  
Of course, we need also to know the strong interaction $BB'M$ vertices, for 
which we utilize
\begin{eqnarray}
{\cal H}_{\rm st}&=&iQ_{\pi NN}\bar{N}\gamma_5\vec{\tau}\cdot
\vec{\pi} N+iQ_{\pi\Sigma\Lambda}
\bar{\vec{\Sigma}}\cdot\vec{\pi}\gamma_5\Lambda\nonumber\\
&+&iQ_{\eta NN}\bar{N}\gamma_5\eta N
+iQ_{\eta\Lambda\Lambda}\bar{\Lambda}\gamma_5\eta\Lambda\nonumber\\
&+&iQ_{KN\Lambda}\bar{N}\gamma_5K\Lambda+iQ_{K\Sigma N}K^\dagger
\bar{\vec{\Sigma}}
\cdot\vec{\tau}\gamma_5N\nonumber\\
&+&iQ_{\rho NN}\bar{N}(\gamma_\mu -i\chi_V{1\over 2M}\sigma_{\mu\nu}q^\nu)
\vec{\rho}^\mu\cdot\vec{\tau} N\nonumber\\
&+&iQ_{\rho\Sigma\Lambda}\bar{\vec{\Sigma}}\cdot\vec{\rho}^\mu
(\gamma_\mu-i\chi_Y{1\over 2M}
\sigma_{\mu\nu}q^\nu)\Lambda\nonumber\\
&+&iQ_{\omega NN}\bar{N}(\gamma_\mu-i\chi_Z{1\over 2M}\sigma_{\mu\nu}q^\nu)
\omega^\mu N\nonumber\\
&+&iQ_{\omega\Lambda\Lambda}\bar{\Lambda}(\gamma_\mu-i\chi_L{1\over 2M}
\sigma_{\mu\nu}q^\nu)\omega^\mu\Lambda\nonumber\\
&+&iQ_{K^*N\Lambda}\bar{N}(\gamma_\mu-i\chi_T{1\over 2M}\sigma_{\mu\nu}q^\nu)
K^{*\mu}\Lambda\nonumber\\
&+&iQ_{K^*\Sigma N}K^{*\mu\dagger}\bar{\vec{\Sigma}}\cdot\vec{\tau}
(\gamma_\mu-i\chi_G
{1\over 2M}\sigma_{\mu\nu}q^\nu)N
\end{eqnarray}
where M is the nucleon mass and $q=p_i-p_f$ is the recoil momentum 
carried off by the meson.  The strong pseudoscalar couplings are determined 
in terms of SU(3) symmetry and the generalized Goldberger-Treiman relation, 
as discussed earlier, yielding
\begin{equation}
\begin{array}{ll}
Q_{\pi NN}={m_N\over F_\pi}(d+f) & Q_{\pi\Sigma\Lambda}={m_\Sigma+m_\Lambda
\over \sqrt{3}F_\pi}d\nonumber\\
Q_{\eta NN}={m_N\over \sqrt{3}F_\pi}(3f-d)& Q_{\eta\Lambda\Lambda}=
-{2m_\Lambda\over \sqrt{3}F_\pi}d\nonumber\\
Q_{KN\Lambda}=-{m_\Lambda+m_N\over 2\sqrt{3}F_\pi}(d+3f)& Q_{K\Sigma N}=
{m_\Sigma+m_N\over 2F_\pi}(d-f)
\end{array}
\end{equation}
with $f,d$ given in Eqn. 30.

The strong vector meson couplings can be found via the 
combined assumptions of SU(3) symmetry and vector dominance, whereby

\begin{equation}
\langle B'|V^i_\mu|B\rangle={m_\rho F_\pi\over m_i^2}\langle \phi^i_\mu B'|
B\rangle
\end{equation}
Putting this together with the nucleon-nucleon matrix element of the 
electromagnetic current
\begin{equation}
\langle N|V^{em}_\mu|N\rangle=\bar{N}[\gamma_\mu{1\over 2}(1+\tau_3)-{i\over 2M}
\sigma_{\mu\nu}q^\nu(1.85\tau_3-0.06)]N
\end{equation}
we have
\begin{eqnarray}
\langle \phi^jB'|B\rangle &=&{-im_j^2\over m_\rho F_\pi}\epsilon^\mu
\bar{B}'[-if_{jB'B}\gamma_\mu\nonumber\\
& -&i(-0.83if_{jB'B}+2.87d_{jB'B})
\sigma_{\mu\nu}q^\nu{1\over 2M}]B
\end{eqnarray}

Finally, we can evaluate the meson exchange diagrams contributing 
to the decay process, yielding for the parity-violating case
\begin{eqnarray}
V^{(-)}(r)&=&{i\over 4}Q_{\pi NN}A_{\pi\Lambda N}(U-3Z)[T(\vec{\sigma}
_\Lambda+
\vec{\sigma}_N)T
-(\vec{\sigma}_\Lambda-\vec{\sigma}_N)]\cdot{\vec f}_\pi^{(-)}(r)\nonumber\\
&+&{i\over 4}Q_{\eta NN}A_{\eta\Lambda N}[T(\vec{\sigma}_\Lambda+
\vec{\sigma}_N)T
-(\vec{\sigma}_\Lambda-\vec{\sigma}_N)]\cdot{\vec f}_\eta^{(-)}(r)\nonumber\\
&-&{i\over 4}Q_{KNN}[(C_{KNN}+D_{KNN})U-(C_{KNN}-D_{KNN})Z]\nonumber\\
&\times&[T(\vec{\sigma}_\Lambda+\vec{\sigma}_N)T-(\vec{\sigma}
_\Lambda-\vec{\sigma}_N)]\cdot{\vec f}_K^{(-)}
(r)\nonumber\\
&+&{i\over 2}Q_{\rho NN}A_{\rho\Lambda N}(U-3Z)\{[T(\vec{\sigma}_\Lambda
+\vec{\sigma}_N)T
+(\vec{\sigma}_\Lambda-\vec{\sigma}_N)]\cdot{\vec f}_\rho^{(+)}(r)\nonumber\\
&+&(1+\chi_V)[T(\vec{\sigma}_\Lambda+\vec{\sigma}_N)S-
(\vec{\sigma}_\Lambda-\vec{\sigma}_N)T]\cdot
{\vec f}_\rho^{(-)}(r)\}\nonumber\\
&+&{i\over 2}Q_{\omega NN}A_{\omega\Lambda N}\{[T(\vec{\sigma}_\Lambda+
\vec{\sigma}_N)T
+(\vec{\sigma}_\Lambda -\vec{\sigma}_N)]\cdot{\vec f}_\omega^{(+)}(r)\nonumber\\
&+&(1+\chi_Z)[T(\vec{\sigma}_\Lambda -\vec{\sigma}_N)S-
(\vec{\sigma}_\Lambda -\vec{\sigma}_N)T]\cdot
\vec{f}_\omega^{(-)}(r)\}\nonumber\\
&+&{i\over 2}Q_{K^*NN}[(C_{K^*NN}+D_{K^*NN})U-(C_{K^*NN}-D_{K^*NN})Z]\nonumber\\
&\times &\{[T(\vec{\sigma}_\Lambda+\vec{\sigma}_N)T-
(\vec{\sigma}_\Lambda-\vec{\sigma}_N)T]\cdot
\vec{f}_{K^*}^{(+)}(r)\nonumber\\
&+&(1+\chi_T)[T\vec{\sigma}_\Lambda-\vec{\sigma}_N)S -
(\vec{\sigma}_\Lambda-\vec{\sigma}_N)T]\cdot
{\vec f}_{K^*}^{(-)}(r)\}
\end{eqnarray}
where T,S are the triplet, singlet spin projection operators
\begin{equation}
T={1\over 4}(3+\vec{\sigma}_N\cdot\vec{\sigma}_\Lambda),
\qquad S+{1\over 4}(1-\vec{\sigma}_N
\cdot\vec{\sigma}_\Lambda)
\end{equation}
while U,Z are the triplet, singlet isospin projection operators
\begin{equation}
U={1\over 4}(3+\vec{\tau}_N\cdot\vec{\tau}_\Lambda),
\qquad Z={1\over 4}(1-\vec{\tau}_N\cdot
\vec{\tau}_\Lambda)
\end{equation}
The radial dependence is given by
\begin{eqnarray}
{\vec f}_M^{(-)}(r)=\left[{i\vec{\nabla}_1-i\vec{\nabla}_2
\over 2M},f_M(r)\right]\nonumber\\
{\vec f}_M^{(+)}(r)=\left\{{i\vec{\nabla}_1-i\vec{\nabla}_2
\over 2M},f_M(r)\right\}
\end{eqnarray}
where $f_M(r)=\exp(-m_Mr)/4\pi r$.

The parity-conserving potential is rather more complicated.  Consider, 
{\it e.g.}, Figure 7 for which
\begin{equation}
V_K^{(+)}(r)={Q_{KN\Lambda}Q_{KN\Lambda}A_{\Lambda N}\over m_N-m_\Lambda}(U-Z)
\{\}_K
\end{equation}
where
\begin{eqnarray}
\{\}_K&=&-{1\over 12}(T-3S)f_K^{[jj]}(r)-{1\over 4}(\sigma^j_\Lambda\sigma^k_N
-{1\over 3}\vec{\sigma}_\Lambda\cdot\vec{\sigma}_N
\delta^{jk})f_K^{[jk]}(r)
\nonumber\\
\mbox{and}\hfill\nonumber\\
f_M^{[jk]}(r)&=&\left[{i\nabla^j_1-i\nabla^j_2\over 2M},
\left[{i\nabla^k_1-i\nabla^k_2\over 2M},f_M(r)\right]\right].
\end{eqnarray}
\begin{figure}
\vspace{2.5in}
\caption{K-exchange diagram contributing to parity-conserving $\Lambda N
\rightarrow NN$.}
\end{figure}
To this must be added the appropriate diagrams for $\pi,\eta$ exchange 
yielding
\begin{eqnarray}
V_{P-tot}^{(+)}(r)&=&V_K^{(+)}(r)+{Q_{KN\Lambda}Q_{KN\Sigma}A_{\Sigma N}\over 
m_N-m_\Sigma}(U-Z)\{\}_K\nonumber\\
&+&{Q_{\pi\Sigma\Lambda}Q_{\pi NN}A_{\Sigma N}\over m_N-m_\Sigma}(U-3Z)
\{\}_\pi\nonumber\\
&+&{Q_{\pi NN}Q_{\pi NN}A_{\Lambda N}\over m_\Lambda-m_N}(U-3Z)\{\}_\pi\nonumber\\
&+&{Q_{\eta NN}Q_{\eta NN}A_{\Lambda N}\over m_\Lambda-m_N}\{\}_\eta\nonumber\\
&+&{Q_{\eta\Lambda\Lambda}Q_{\eta NN}A_{\Lambda N}\over m_N-m_\Lambda}\{\}_\eta
\end{eqnarray}

Also we must consider vector exchange.  From Figure 8 we have
\begin{eqnarray}
V_\rho^{(+)}(r)&=&{Q_{\rho\Sigma\Lambda}Q_{\rho NN}A_{\Sigma N}\over 
m_N-m_\Sigma}(U-3Z)F_\rho(V,Y)\nonumber\\
\mbox{with}\hfill\nonumber\\
F_\rho(V,Y)&=&f_\rho(r)+f_\rho(r){1\over 2}\left({i\vec{\nabla}_1-
i\vec{\nabla}_2\over 2M}
\right)^2+{1\over 2}{\vec f}_\rho^{(-)}(r)\cdot{i\vec{\nabla}_1
-i\vec{\nabla}_2\over 2M}
\nonumber\\
&-&{1\over 4}(1+\chi_Y+\chi_V)f_\rho^{[jj]}(r)-{1\over 6}(1+\chi_Y)(1+\chi_V)
f_\rho^{[jj]}(r)(T-3S)\nonumber\\
&+&{1\over 4}(1+\chi_Y)(1+\chi_V)(\sigma^j_\Lambda\sigma^k_N-{1\over 3}
\vec{\sigma}_\Lambda\cdot\vec{\sigma}_N\delta^{jk})f_\rho^{[jk]}(r)T\nonumber\\
&+&{i\over 2}[(1+{3\over 2}\chi_Y)\sigma_\Lambda^n
+(1+{3\over 2}\chi_V)\sigma_N^n]\epsilon_{\ell mn}f^{[\ell]}_\rho(r){i\nabla_1^m
-i\nabla_2^m\over 2M}
\end{eqnarray}
where
\begin{equation}
f_M^{[j]}(r)=\left[{i\nabla_1^j-i\nabla_2^j\over 2M}, f_M(r)\right]
\end{equation}
\begin{figure}
\vspace{2.5in}
\caption{$\rho$-exchange contribution to parity-conserving $\Lambda N
\rightarrow NN$.}
\end{figure}
Again to this must be added the various $\omega,K^*$ exchange terms.  Then
\begin{eqnarray}
V_{V-tot}^{(+)}(r)&=&V_\rho^{(+)}(r)+{Q_{\rho NN}Q_{\rho NN}A_{\Lambda N}\over m_\Lambda
-m_N}(U-3Z)F_\rho(V,V)\nonumber\\
&+&{Q_{K^*N\Lambda}Q_{K^*\Sigma N}A_{\Sigma N}\over m_N-m_\Sigma}(U-Z)
F_{K^*}(T,G)\nonumber\\
&+&{Q_{K^*N\Lambda}Q_{K^*N\Lambda}A_{\Lambda N}\over m_N-m_\Lambda}(U-Z)
F_{K^*}(T,T)\nonumber\\
&+&{Q_{\omega NN}Q_{\omega NN}A_{\Lambda N}\over m_\Lambda-m_N}F_\omega(Z,Z)
\nonumber\\
&+&{Q_{\omega \Lambda\Lambda}Q_{\omega NN}A_{\Lambda N}\over m_N-m_\Lambda}
F_\omega(L,Z)
\end{eqnarray}
We must also consider the various double pole diagrams.  For the $K\pi$
term we have
\begin{eqnarray}
V_{K\pi}(r)&=&-{1\over 4}{Q_{\pi NN}Q_{KN\Lambda}A_{\pi K}\over m_K^2-m_\pi^2}
(U-3Z)\sigma_\Lambda^j\sigma_N^k\nonumber\\
&\times &\left[{i\nabla_1^k-i\nabla_2^k\over 2M},\left[{i\nabla^j_1-i\nabla^j_2
\over 2M}, f_K(r)-f_\pi(r)\right]\right]
\end{eqnarray}
to which must be added the $V_{K\eta}$ pieces
\begin{eqnarray}
V_{K\eta}(r)&=&-{1\over 4}{Q_{KN\Lambda}Q_{\eta NN}A_{\eta K}\over m_K^2-m_\pi^2}
\sigma_\Lambda^j\sigma_N^k\nonumber\\
&\times &\left[{i\nabla_1^k-i\nabla_2^k\over 2M},\left[{i\nabla_1^j-i\nabla_2^j
\over 2M},f_K(r)-f_\eta(r)\right]\right]
\end{eqnarray}
Finally we must append the $K^*\rho $ and $K^*\omega$ potentials:
\begin{eqnarray}
V_{K^*\rho}(r)+V_{K^*\omega}(r)&=&{Q_{K^*N\Lambda}Q_{\rho NN}A_{\rho K^*}
\over m_{K^*}^2
-m_\rho^2}(U-3Z)[F_{K^*}(T,V)-F_\rho(T,V)]\nonumber\\
&+&{Q_{K^*N\Lambda}Q_{\omega NN}A_{\omega K^*}\over m_{K^*}^2-m_\omega^2}
(U-3Z)[F_{K^*}(T,Z)-F_\omega(T,Z)].\nonumber\\
\qquad
\end{eqnarray}

Having thus constructed the $\Lambda N$ ``potentials" which give rise to 
hypernuclear decays, the task remaining is to evaluate their matrix elements 
in the nuclear medium, as will be described in the next section.

\section{Hypernuclear Decay: Nuclear Matter Calculations}

An ideal calculation of the decay of $5\leq A\leq 40$ hypernuclei would proceed 
using shell-model wavefunctions for the specific nucleus under consideration.  
However, as a first step in such a program we present here a calculation in 
so-called ``nuclear matter".  This is useful both as an approximation to the
realistic decay of heavy hypernuclei and as a calibration with previous work
in the field.

The calculation is performed utilizing the potentials given in Eqns. 49-61.
A Fermi gas model is assumed with $N_n=N_p$ and $k_F=270$ MeV.
We include in the initial state only the relative S-wave projection of the
$\Lambda$ with each of the A nucleons---indeed the $\Lambda$ is very 
weakly bound and its wavefunction is correspondingly quite diffuse.  The decay 
rate is then given in terms of the expression for the $\Lambda$ interaction 
taking place at rest
\begin{equation}
\Gamma_{NM}={1\over (2\pi)^5}\int_0^{k_F}d^3k\int d^3k_1\int d^3k_2
\delta^4(p_\Lambda +k-k_1-k_2){1\over 4}\sum_{s_i,s_f} |\langle f|V|i
\rangle |^2
\end{equation}
which can be reduced to
\begin{equation}
\Gamma_{NM}={1\over 8\pi}\left({m_\Lambda+m_N\over m_\Lambda}\right)^3
\int_0^{k_F({m_\Lambda\over m_\Lambda+m_N)}}p^2dp\sum_{\beta\alpha}R(
\beta\leftarrow\alpha)
\end{equation}
where
\begin{equation}
R(\beta\leftarrow\alpha)=qm_N|\langle\beta|V|\alpha\rangle |^2
\end{equation}
with
\begin{equation}
q^2=m_N(m_\Lambda-m_N)+{p^2\over 2\mu_{\Lambda N}}
\end{equation}
are the basic transition rates between the various $\Lambda$N and NN 
configurations---$\alpha$ and $\beta$---and $p,q$ are the relative 
$\Lambda$N, NN momenta respectively.\footnote{Here the sum over alpha and beta
involves the various spin and isospin indices.  Our notation below differs
from that of other authors in that we subsume spin and isospin factors into
the partial rates;  details are given by B. Gibson\cite{30}.}
There is at least one subtlety which we need to emphasize at this point 
having to do with the generically written Yukawa functions
\begin{equation}
f_M(r)={e^{-m_Mr}\over 4\pi r}
\end{equation}
which pervade the potential forms given earlier.  The origin of these 
terms is the Fourier transform of the usual momentum space propagator
\begin{equation}
f_M(r)=\int {d^3q\over (2\pi)^3}e^{i{\vec q}\cdot{\vec r}}{1\over m_M^2+
{\vec q}^2 -q_0^2}
\end{equation}
For a typical nucleon-nucleon interaction, we have
\begin{eqnarray}
q_0&=&m_N+{p_1^2\over 2m_N}-m_N-{p_2^2\over 2m_N}\nonumber\\
&=&
{(p_1+p_2)|{\vec q}|\over 2m_N}< {p_F\over m_N}|{\vec q}|<< 
|{\vec q}| 
\end{eqnarray}
so that
\begin{equation}
f_M(r)=\int{d^3q\over (2\pi)^3}e^{i{\vec q}\cdot{\vec r}}
{1\over m_M^2+{\vec q}^2}
={e^{-m_Mr}\over 4\pi r}
\end{equation}
as given above.  However, if the meson connects to a $\Lambda$N vertex 
then for a $\Lambda$ at rest
\begin{equation}
q_0\approx{1\over 2}(m_\Lambda-m_N)
\end{equation}
The Fourier transform then becomes
\begin{equation}
\tilde{f}_M(r)=\int {d^3q\over (2\pi)^3}e^{i{\vec q}\cdot{\vec r}}
{1\over m_M^2+{\vec q}^2-{1\over 4}(m_\Lambda-m_N)^2}=
{e^{-\tilde{m}_Mr}\over 4\pi r}
\end{equation}
where
\begin{equation}
\tilde{m}_M=\sqrt{m_M^2-{1\over 4}(m_\Lambda-m_N)^2}
\end{equation}
The important feature here is that the effective range of the exchange 
potential is {\it increased}, since $\tilde{m}_M < m_M$.  Thus we have
\begin{equation}
\begin{array}{cc}
 meson & \tilde{m}_M / m_M \\ 
\pi & 0.76 \\
K & 0.98 \\
\rho & 0.99 
\end{array}
\end{equation}
so that the effect is important for pions and is understood to be included 
in the evaluation of all the potential expressions in Eqns. 49-61.

At this point, it is interesting to perform a preliminary calculation.  Since 
the $\Lambda$ is taken to be in a relative S-state with respect to any of the
core nucleons, the initial ${}^{2S+1}L_J$ configuration must be either 
${}^1S_0$ or ${}^3S_1$.  As the weak interaction can either change or not
change the parity, there exist six possible transitions $\alpha\rightarrow
\beta$
\begin{eqnarray}
{}^1S_0\rightarrow\left\{\begin{array}{c}
{}^3P_0 \\
{}^1S_0
\end{array}\right.\nonumber\\
{}^3S_1\rightarrow\left\{\begin{array}{c}
{}^3P_1\\
{}^1P_1\\
{}^3S_1\\
{}^3D_1
\end{array}\right.
\end{eqnarray}

We first perform the calculation for a simple pion-only-exchange 
potential, with the weak $\Lambda N\pi$ interaction given in Eqn. 2,
which matches the on shell weak couplings observed experimentally.  It 
is useful to characterize the various allowed $\Lambda N\rightarrow NN$ 
modes in terms of effective transition operators, as given in Table 2.
\begin{table}
\begin{center}
\begin{tabular}{|l|l|}\hline
Transition & Operator \\ 
\hline
\hline
${}^1S_0\rightarrow{}^1S_0(I=1)$ & ${1\over 4}a(q^2)
(1-\vec{\sigma}_\Lambda\cdot\vec{\sigma}_N)$\\ \hline
${}^1S_0\rightarrow{}^3P_0(I=1)$ &$ {1\over 8}b(q^2)(\vec{\sigma}_\Lambda
-\vec{\sigma}_N)\cdot\hat{q}(1-\vec{\sigma}_\Lambda\cdot
\vec{\sigma}_N)$\\ \hline
${}^3S_1\rightarrow{}^3S_1(I=0)$ &$ {1\over
4}c(q^2)(3+\vec{\sigma}_\Lambda\cdot
\vec{\sigma}_N)$\\
\hline
${}^3S_1\rightarrow{}^3D_1(I=0)$ &$ {3\over 2\sqrt{2}}d(q^2)
(\vec{\sigma}_\Lambda\cdot\hat{q}\vec{\sigma}_N
\cdot\hat{q}-{1\over 3}\vec{\sigma}_\Lambda\cdot\vec{\sigma}_N)$ \\ \hline
${}^3S_1\rightarrow{}^1P_1(I=0)$ &$ {\sqrt{3}\over 8}e(q^2)(\vec{\sigma}
_\Lambda 
-\vec{\sigma}_N)\cdot\hat{q}(3+\vec{\sigma}_\Lambda\cdot
\vec{\sigma}_N)$ \\ \hline
${}^3S_1\rightarrow{}^3P_1(I=1)$ &$ {\sqrt{6}\over 4}f(q^2)(\vec{\sigma}
_\Lambda
+\vec{\sigma}_N)\cdot\hat{q}$ \\ \hline
\end{tabular}
\end{center}
\caption{Transition operators of allowed $\Lambda N\rightarrow NN$ transitions
from relative S-states.  Here $\vec{q}$ specifies the relative momentum of the 
outgoing nucleons while $\vec{\sigma}_\Lambda,\vec{\sigma}_N$ 
operate on the $\Lambda N,NN$ vertices respectively.}
\end{table}
We find then for the total nonmesonic hypernuclear decay rate
\begin{equation}
\Gamma_{NM}={3\over 8\pi\mu^3_{\Lambda N}}\int_0^{\mu_{\Lambda N}k_F}
dpp^2m_Nq\left(|a|^2+|b|^2+|c|^2+|d|^2+|e|^2+3|f|^2\right)
\end{equation}
where $\mu_{\Lambda N}={m_\Lambda\over m_\Lambda+m_N}$ arises from the switch 
from the nuclear rest frame to the $\Lambda N$ center of mass frame. 
The results of the calculation are given in Table 3.
\begin{table}
\begin{center}
\begin{tabular}{|c|c|}\hline
channel & $\Gamma_{NM}/\Gamma_\Lambda$\\ 
\hline
\hline
${}^1S_0\rightarrow{}^1S_0$ & 0.381\\ \hline
${}^1S_0\rightarrow{}^3P_0$ & 0.160\\ \hline
${}^3S_0\rightarrow{}^3P_1$& 0.319\\ \hline
${}^3S_1\rightarrow{}^1P_1$ & 0.478\\ \hline 
${}^3S_1\rightarrow{}^3S_1$ & 0.381\\ \hline
${}^3S_1\rightarrow{}^3D_1$ & 2.94 \\ \hline
\end{tabular}
\end{center}
\caption{Hypernuclear decay rates in nuclear matter for pion-only-exchange and 
no correlations.}
\end{table}
Note that about 20\% of the transition rate comes from the parity violating 
sector and that 65\% comes from the single branch ${}^3S_1\rightarrow{}^3D_1$!

A comparison with other pion-only-exchange calculations of hypernuclear 
decay in nuclear matter is provided in the first line of Table 4.
\begin{table}
\begin{center}
\begin{tabular}{|c|c|c|c|c|}\hline
   & Adams\cite{13} & McK-Gib\cite{28} & Oset-Sal\cite{29} & our calc. \\
\hline
\hline
${1\over \Gamma_\Lambda}\Gamma_{NM}$(no corr.)& 0.51 & 4.13 &4.3 & 4.66\\
\hline
${1\over \Gamma_\Lambda}\Gamma_{NM}$(corr.) & 0.06 & 2.31 & 2.1 & 1.85 \\ \hline
\end{tabular}
\end{center}
\caption{Non-mesonic hypernuclear decay rates calculated by various groups
using pion-exchange only in ``nuclear matter."}
\end{table}
We note that there exists general agreement---$(\Gamma_{NM}/\Gamma_\Lambda)
_\pi\sim 4.5$---except for the calculation of Adams, whose result is nearly an
order of magnitude smaller.   This is primarily due to his incorrect use of a 
$\Lambda N\pi$ coupling which is a factor of three too small
\begin{equation}
g_w^{\rm Adams}=9.0\times 10^{-8}\quad{\rm vs.}\quad g_w^{\rm exp}=2.35\times
10^{-7}.
\end{equation}
When the correct value is used, Adams result becomes $(\Gamma_{NM}/\Gamma_
\Lambda)=3.6$ and agrees with the other calculations of the hypernuclear decay
rate.

Our preliminary calculation is clearly naive in that it neglects both short 
range correlations and the effects of additional meson exchanges.  One 
indication of the importance of the former can be seen from a simple argument.  
Consider a scalar exchange between nucleons in relative S-states.  
We have then
\begin{equation}
\langle f(L=0)|V|i(L=0)\rangle\sim \mu^2\int\psi_f^*\psi_i{e^{-\mu r}\over 
4\pi r}d^3r
-\int\psi_f^*\psi_i\delta^3(r)d^3r
\end{equation}
Now drop the $\delta^3(r)$ component---indeed any short range correlation 
which gives $\psi_i(r=0)=0$ or $\psi_f(r=0)=0$ will eliminate $S\rightarrow S$
transitions essentially entirely, since for $\mu\sim m_\pi$
\begin{equation}
\int\psi_f^*\psi_i\delta^3(r)d^3r\sim 80\mu^2\int\psi_f^*\psi_i
{e^{-\mu r}\over 4\pi r}d^3r
\end{equation}
This modification changes our result in Table 4 to
\begin{equation}
{\Gamma_{NM}\over \Gamma_\Lambda}(\mbox{no}\, S\rightarrow S)=2.18
\end{equation}
which is a considerable shift and emphasizes the need to consistently
include the effects of initial state correlations.

Other authors have also been concerned about such correlation effects.  
Adams in his work used for the initial state correlation a hard core of
$r_c=0.4$ fm with a solution of the Bethe-Goldstone equation
\begin{equation}
f(r)\sim j_0(qr)-{\sin(qr_c){\rm si}(\beta r)\over qr{\rm si}(\beta r_c)}
\end{equation}
employed for $r > r_c$.  For the final nucleon-nucleon state he utilized
a square well potential which also contained a hard core.  An unusually strong 
tensor correlation was used.  On the other hand McKellar and Gibson employed 
an effective form for the initial state correlation
\begin{equation}
f(r)=1-\exp(-\alpha r^2)
\qquad\mbox{with}\qquad \alpha=1.8\,\mbox{fm}^{-2}
\end{equation}
and a Ried soft-core potential for the final state correlation.  Likewise
below we shall utilize an
initial state correlation of the
McKellar-Gibson form while for the final state we use a Reid soft-core 
potential to generate the final state interaction.  Performing the 
calculation with either or both correlations included, we find the results 
given in Table 5.
\begin{table}
\begin{center}
\begin{tabular}{|c|c|c|c|c|}\hline
\quad & no corr. & init. st. only & fin. st. only & 
both corr. \\ 
\hline
\hline
${}^1S_0\rightarrow{}^1S_0$ & 0.381 & 0.0002 & 0.0007 & 0.0002 \\ \hline
${}^1S_0\rightarrow{}^3P_0$ & 0.160 & 0.070 & 0.048 & 0.037 \\ \hline
${}^3S_1\rightarrow{}^3P_1$ & 0.319 & 0.140 & 0.180 & 0.119 \\ \hline 
${}^3S_1\rightarrow{}^1P_1$ & 0.478 & 0.209 & 0.167 & 0.130 \\ \hline
${}^3S_1\rightarrow{}^3S_1/{}^3D_1$ & 0.381 & 0.0002 & 2.02 & 0.797 \\ \hline
${}^3S_1\rightarrow{}^3S_1/{}^3D_1$ & 2.94 & 1.76 & 0.838 & 0.760 \\ \hline
Total & 4.66 & 2.18 & 3.25 & 1.85 \\ \hline
\end{tabular}
\end{center}
\caption{$\Gamma_{NM}/\Gamma_\Lambda$ in nuclear matter for pion-only-exchange
and including the effects of correlations.}
\end{table}
Note the almost complete suppression of $S\rightarrow S$ transitions (with 
final state tensor interactions included, the effect is obscured by the mixing of
${}^3S_1$ and ${}^3D_1$ states.)

The comparison with other calculations, including correlations is given in 
line two of Table 4.  We see again that there is general 
agreement---$(\Gamma_{NM}/\Gamma_\Lambda)_\pi^{\rm corr.}\sim 2.0$---except for
that of Adams, for which the origin of the discrepancy is now twofold.  The
first problem is the use of an incorrect weak coupling, as mentioned earlier.
The other is the use of an inappropriately strong tensor correlation, which 
suppresses the very important ${}^3S_1\rightarrow{}^3D_1$ transition.  When
both effects are corrected Adams' number becomes $(\Gamma_{NM}/\Gamma_\Lambda)_\pi
^{\rm corr.}\sim 1.7$ and is in agreement with other authors.

From this initial pion-exchange-only 
nuclear matter calculation then we learn that the basic 
nonmesonic decay rate is anticipated to be of the same order as that 
for the free lambda {\it and} the
important role played by correlations.  A second quantity of interest which 
emerges 
from such a calculation is the p/n stimulated decay ratio, which has been
calculated by two of the groups, results of which are displayed in Table
6.
\begin{table}
\begin{center}
\begin{tabular}{|c|c|c|c|c|}\hline
    & Adams\cite{13} & McK-Gib\cite{28} & Oset-Sal\cite{29} & our calc. \\ 
\hline
\hline
$\Gamma_{NM}(p/n)$ (no corr.) & 19.4 & - & -& 11.2 \\ \hline
$\Gamma_{NM}(p/n)$ (corr.) & 2.8  & - & - & 16.6 \\ \hline
\end{tabular}
\end{center}
\caption{Proton to neutron stimulated decay ratios for pion-only exchange in
``nuclear matter."}
\end{table}
An interesting feature here is that the numbers come out so large---proton
stimulated decay is predicted to predominate over its 
neutron stimulated counterpart by nearly an order of magnitude.
The reason for this is easy to see.\cite{30}  In a pion-exchange-only scenario the
effective weak interaction is of the form
\begin{equation}
{\cal H}_w\sim g\bar{N}\vec{\tau} N\cdot \bar{N}\vec{\tau} \Lambda
\end{equation}
Then $\Lambda n\rightarrow nn\sim g$ but $\Lambda p\rightarrow np\sim (-1-
(\sqrt{2})^2)g=-3g$ since both charged and neutral pion exchange are involved.
In this naive picture then we have $\Gamma_{NM}(p/n)\sim 9$, in rough
agreement with the numbers given in Table 6.

Armed finally with theoretical expectations, we can ask what does experiment 
say?  The only reasonably precise results obtained for nuclei with $A> 4$ 
were those measured at BNL and summarized in Table 1.  
We observe that the measured nonmesonic decay rate is about a factor of two
lower than that predicted in Table 4 while the p/n stimulation ratio differs
by at least an order of magnitude from that given in Table 6.  The problem
may be, of course, associated with the difference between the nuclear
matter within which the calculations were performed and the finite
nuclear systems which were examined experimentally.  Or it could be due to
the omission of the many shorter range exchanged mesons in the theoretical
estimate. (Or {\it both!})  

Before undertaking the difficult problem of 
finite nuclear calculations, it is useful to first examine the inclusion of 
additional
exchanged mesons in our calculations.  As mentioned above, a primary difficulty
in this approach 
is that none of the required weak couplings can be measured experimentally.
Thus the use of some sort of model is required, and the significance of any
theoretical predictions will be no better than the validity of the model.
One early attempt by McKellar and Gibson,\cite{28} for
example, included only the rho and evaluated the rho couplings using SU(6) and 
alternatively the straightforward but flawed factorization approach.
Well aware of the limitations of
the latter they allowed an arbitrary phase between the rho and pi amplitudes
and renormalized the factorization calculation by a factor of
$1/\sin\theta_c\cos\theta_c$ in order to account for the $\Delta I={1\over 2}$
enhancement.  Obviously this is only a rough estimate then and this is only for the
rho meson exchange contribution!  A similar approach was attempted by 
Nardulli, who calculated the parity-conserving rho amplitude in a simple 
pole model and the parity violating piece in a simple quark picture.\cite{31}  
Results of these calculations are shown in Table 7.
\begin{table}
\begin{center}
\begin{tabular}{|c|c|c|c|}\hline
    & McK-Gib\cite{28} $\pi+\rho$ & McK-Gib\cite{28} $\pi-\rho$ & 
Nard.\cite{30} \\ 
\hline
\hline
${1\over \Gamma_\Lambda}\Gamma_{NM}$ & 3.52 & 0.72 & 0.7 \\ \hline
\end{tabular}
\end{center}
\caption{Nonmesonic decay rates in nuclear matter in pi plus rho exchange models}
\end{table}

To our knowledge, the only comprehensive calculation which has been undertaken 
to date is that of our group, which is described in the previous section of this
paper.  We show in Table 8 the results of including the effects of additional
meson exchange contributions in arbitrary order.
\begin{table}
\begin{center}
\begin{tabular}{|c|c|c|c|}\hline
Exchanged Meson & S$\rightarrow$P & S$\rightarrow$S/D & Total \\ 
\hline
\hline
$\pi$ & 0.288 & 1.56 & 1.85 \\ \hline
$+\eta$  &0.320 & 1.36 & 1.68 \\ \hline
$+K$ & 0.568 & 0.74 & 1.31 \\ \hline
$+\rho$ & 0.523 & 0.59 & 1.11 \\ \hline
$+\omega$ & 0.576 & 0.61& 1.19 \\ \hline
$+K^*$ & 0.721 & 0.66 & 1.38 \\ \hline
\end{tabular}
\end{center}
\caption{Hypernuclear decay rates---$\Gamma_{NM}/\Gamma_\Lambda$---in nuclear
matter including the effects of correlations and with the contributions of 
non-pion exchanges.}
\end{table}
We see from Table 8 that inclusion of {\it all} pseudoscalar and vector meson
exchanges in addition to the long-range pion-exchange component has the effect
of reducing the hypernuclear decay rate to a value about 40\% above that for
free $\Lambda$-decay and in agreement with the $A\sim 12$ results.  However, 
in view of the theoretical uncertainties associated with our calculation and
the fact that it is performed in nuclear matter, this agreement cannot be
said to distinguish between the pion-only-exchange and all-exchanges scenarios.

A more convincing case for the presence of non-pion-exchange {\it can} be
constructed by examining the proton/neutron stimulation ratio.  On the 
theoretical side, we find
\begin{equation}
\Gamma_{NM}(p/n)={\int_0^{\mu_{\Lambda N}k_F}p^2dpq(|a|^2+|b|^2+3|c|^2+3|d|^2
+3|e|^2+3|f|^2)\over \int_0^{\mu_{\Lambda}k_F}p^2dpq(2|a|^2+2|b|^2+6|f|^2)}
\end{equation}
in terms of the couplings $a,b,...f$ defined in Table 2.  
Results are shown in Table 9.  
\begin{table}
\begin{center}
\begin{tabular}{|c|c|c|}\hline
    &$ \Gamma_{NM}(p/n)$ & $\Gamma_{NM}(PV/PC)$\\ 
\hline
\hline
$\pi$ (no corr.) &  11.2 & 0.14 \\ \hline
$\pi$ (with corr.) & 16.6 & 0.18 \\ \hline
$\pi+\rho $  & 13.1 & 0.21 \\ \hline
$\pi,\rho,\omega,\eta,K,K^*$ & 2.9 & 0.90 \\ \hline
\end{tabular}
\end{center}
\caption{The parity violating to parity conserving and p to n ratios for
hypernuclear decay in ``nuclear matter."}
\end{table}
We observe that inclusion of additional exchanges plays a major role in reducing 
the p/n ratio from its pion-only-exchange value.  The resulting value of 
2.9 is still somewhat larger than the experimental numbers shown in Table 1
but clearly indicates the presence of non-pion-exchange components.

The reason that kaon exchange in particular can play such a major role 
can be seen from a simple argument due to Gibson\cite{30} who pointed out 
that since the final NN system can have either I=0 or I=1, the effective
kaon exchange interaction can be written as
\begin{eqnarray}
{\cal L}_{eff}=A_0(\bar{p}p+\bar{n}n)\bar{n}\Lambda+A_1(2\bar{n}p\bar{p}\Lambda
-(\bar{p}p-\bar{n}n)\bar{n}\Lambda)\nonumber\\
\sim(A_0-3A_1)\bar{p}p\bar{n}\Lambda+(A_0+A_1)\bar{n}n\bar{n}\Lambda
\end{eqnarray}
where the second line is obtained via a Fierz transformation.  Since for
parity-violating kaon exchange we have $A_0\sim 6.8A_1$\footnote{From Eqs.
34,40 we determine
\begin{equation}
{A_0\over A_1}= {C_{KNN}+2D_{KNN}\over C_{KNN}}=3(1-{b_V\over c_V})\simeq
6.8\nonumber\\
\end{equation}} we find\cite{30}
\begin{equation}
\Gamma_{NM}(p/n)=\left({A_0-3A_1\over A_0+A_1}\right)^2\sim 1/4
\end{equation}
which clearly indicates the importance of inclusion of non-pion-exchange 
components in predicting the p/n ratio.

A second strong indication of the presence of non-pion-exchange can 
be seen from Table 9 in that the ratio of rates for parity-violating to 
parity-conserving transitions is substantially enhanced by the inclusion of 
kaon and vector meson exchange as compared to the simple pion-exchange-only 
calculation.  We can further quantify this effect by calculating explicitly 
the angular distribution of the emitted proton in the $\Lambda p\rightarrow np$
transition (there can be no asymmetry for the corresponding 
$\Lambda n\rightarrow nn$ case due to the identity of the final state neutrons),
yielding
\begin{equation}
W_p(\theta)\sim 1+\alpha P_\Lambda\cos\theta
\end{equation}
where
\begin{equation}
\alpha={\int_0^{\mu_{\Lambda N}k_F}p^2dpq{\sqrt{3}\over 2}{\rm Re}f^*
(\sqrt{2}c+d)\over \int_0^{\mu_{\Lambda N}k_F}p^2dpq{1\over 4}(|a|^2+|b|^2
+3|c|^2+3|d|^2+3|e|^2+3|f|^2}
\end{equation}
is the asymmetry parameter.  Results of a numerical evaluation are shown in
Table 10
\begin{table}
\begin{center}
\begin{tabular}{|c|c|c|c|}\hline
\quad & $\pi$-no corr. & $\pi$-corr. & all exch. \\
\hline
\hline
$\alpha$ & -0.078 & -0.192 & -0.443 \\ \hline
\end{tabular}
\end{center}
\caption{Proton asymmetry coefficient in various scenarios.}
\end{table}
so that again inclusion of non-pion-exchange components has a significant 
effect, increasing the expected $\Lambda p\rightarrow np$ asymmetry 
by more than a factor of two.  This prediction of a substantial asymmetry
is consistent with preliminary results obtained for p-shell nuclei at 
KEK.\cite{32}

\section{Hypernuclear Decay: Additional Considerations}

Although nuclear matter calculations are of great interest in identifying
basic properties of the decay process, true confrontation with experiment
requires calculations involving the finite nuclei on which the 
measurements are conducted.  Of course, such calculations are 
considerably more demanding than their nuclear matter counterparts and 
require $\Lambda$ shell model considerations as well as non-S-shell capture.  
Nevertheless a number of groups have taken up the challenge.  Details of 
our own calculation in ${}^{12}_\Lambda$C and ${}^5_\Lambda$He will be
presented in a future publication.  These are performed using a simple 
shell model to describe the hypernuclear structure, where only an extreme
single particle model with no configuration mixing and only 
phenomenological forms for the correlation functions are employed.  In 
Table 11 we compare our preliminary results for ${}^{12}_\Lambda$C with 
that obtained in a parallel calculation performed by a TRIUMF 
collaboration\cite{33} and with a pion-only-exchange version by Oset 
and Salcedo.\cite{29}
\begin{table}
\begin{center}
\begin{tabular}{|c|c|c|c|}\hline
  & Oset-Sal\cite{29}& TRIUMF\cite{33} & our calc. \\ 
\hline
\hline
${1\over \Gamma_\Lambda}\Gamma_{NM} \pi$ (no corr.)&  & 1.6 & 3.4 \\ \hline 
$\qquad\pi$ (corr.) & 1.5 & 2.0(1.0) & 0.5\\ \hline
$\pi+K\cite{33};\,\,\,\pi,\eta,\rho\omega,K,K^*\cite{34}$& & 1.2 & 0.2 \\ \hline
$\Gamma_{NM}(p/n) \pi$ (no corr.)& & 5.0 & 4.6 \\ \hline
\qquad $\pi$ (corr.) & & 5.0 & 5.0 \\ \hline
$\pi+K\cite{33};\,\,\,\pi,\eta,\rho,\omega,K,K^*\cite{34}$& & 4.0 & 1.2 \\ \hline
$\Gamma_{NM}(PV/PC) \pi$ (no corr.) & & 0.4 & 0.1 \\ \hline
\qquad $\pi$ (corr.) & & 0.5 & 0.1 \\ \hline
$\pi+K\cite{15};\,\,\,\pi,\eta,\rho,\omega,K,K^*\cite{34}$& & 0.3 & 0.8 \\ \hline
\end{tabular}
\end{center}
\caption{Calculated properties of nonmesonic hypernuclear decay of 
${}^{12}_\Lambda$C.}
\end{table}
In comparing with the experimental results given in Table 3, we see that 
our calculation is certainly satisfactory, but the discrepancy with 
the TRIUMF work is disturbing and needs to be rectified before either
is to be believed.

A second nucleus on which there has been a good deal of work, both 
experimentally and theoretically, is ${}^5_\Lambda$He, which is summarized 
in Table 12.
\begin{table}
\begin{center}
\begin{tabular}{|c|c|c|c|c|}\hline
    & Oset-Sal\cite{29}&TRIUMF\cite{33}&TTB\cite{35}&our calc.\\ 
\hline
\hline
${1\over \Gamma_\Lambda}\Gamma_{NM} \pi$(no corr.)&  & 1.0 & 0.5 & 1.6 \\ \hline
$\pi$(corr.) & 1.15 & 0.25(0.5) & 0.144 & 0.9 \\ \hline
$\pi+K\cite{33};\,\,\,\pi,\eta,\rho,\omega,K,K^*\cite{34}$& & 0.22 & & 0.5\\ \hline
$\Gamma_{NM}(p/n) \pi$ (no corr.)&  & 5.0 & & 15 \\ \hline
$\pi$ (corr.)& & 4.8& & 19 \\ \hline
$\pi+K\cite{33};\,\,\,\pi,\eta,\rho,\omega,K,K^*\cite{34}$& & 5.4 & & 2.1\\ \hline
\end{tabular}
\end{center}
\caption{Calculated properties of the nonmesonic decay of ${}^5_\Lambda$He.}
\end{table}
Here again what is important is not so much the agreement or disagreement 
with experiment but rather the discrepancies {\it between} the various 
calculations which need to be clarified before any significant confrontation 
between theory and experiment is possible.  One step in this regard has 
already been taken in that the TRIUMF collaboration have reexamined their
work and have discovered an inconsistency between their calculational
technique and the correlations which were utilized.\cite{36}  When the correlations
are properly included the substantial reductions in the calculated finite
hypernuclear decay rates given in Table 12 are replaced by values which 
are much more in accord with our calculation, as shown in parentheses.
Only the pion exchange piece has been included and there is clearly much
room for additional work.

Before leaving this section, however, it is important to raise an
additional issue which must be resolved before reliable
theoretical calculations are possible---that of the $\Delta I={1\over 2}$
rule.\cite{37}   Certainly in any venue in which it has been tested---nonleptonic kaon
decay---$K\rightarrow 2\pi,3\pi$, hyperon decay---$B\rightarrow B'\pi$,
$\Delta I={1\over 2}$ components of the decay amplitude are found to 
be a factor of twenty or so larger than their $\Delta I={3\over 2}$ 
counterparts.  Thus it has been natural in theoretical analysis of nonmesonic
hypernuclear decay to make this same assumption.  (Indeed without it the 
already large number of unknown parameters in the weak vertices expands 
by a factor of two.)  However, recently Schumacher has raised a serious
question about the correctness of this assumption, which if verified will
have serious implications about the direction of future theoretical
analyses.\cite{38}  The point is that by use of very light hypernuclear systems
one can isolate the isospin structure of the weak transition.  Specifically,
using a simple delta function interaction model of the hypernuclear
weak decay process, as first written down by Block and Dalitz in 1963, one 
determines\cite{39}
\begin{eqnarray}
{}^4_\Lambda \mbox{He}: \gamma_4=\Gamma_{NM}(n/p)={2R_{n0}\over 3R_{p1}+R_{p0}}
\nonumber\\
{}^5_\Lambda \mbox{He}: \gamma_5=\Gamma_{NM}(n/p)={3R_{n1}+R_{n0}\over 3R_{p1}+R_{p0}}
\nonumber\\
\gamma=\Gamma_{NM}({}^4_\Lambda \mbox{He})/\Gamma_{NM}({}^4_\Lambda \mbox{H})=
{3R_{p1}+R_{p0}+2R_{n0}\over 3R_{n1}+R_{n0}+2R_{p0}}
\end{eqnarray}
where here $R_{Nj}$ indicates the rate for N-stimulated hypernuclear decay
from an initial configuration having spin j.  One can then isolate the ratio
$R_{n0}/R_{p0}$ by taking the algebraic combination
\begin{equation}
{R_{n0}\over R_{p0}}={\gamma\gamma_4\over 1+\gamma_4-\gamma\gamma_5}
\end{equation}
and from the experimental values\cite{40}
\begin{equation}
\gamma_4=0.27\pm 0.14,\qquad \gamma_5=0.93\pm 0.55,\qquad 
\gamma=0.73^{+0.71}_{0.22}
\end{equation}
one determines
\begin{equation}
{R_{n0}\over R_{p0}}={0.20^{+0.22}_{-0.12}\over 0.59^{+0.80}_{-0.47}}
\end{equation}
in possible conflict with the $\Delta I={1\over 2}$ rule 
prediction---$R_{n0}/R_{p0}=2$.\footnote{Note that final state nn or np 
configurations which arise from initial ${}^1S_0$ states are of necessity 
I=1.}  If confirmed by further theoretical and 
experimental analysis this would obviously have important ramifications for
hypernuclear predictions.  However, recent work at KEK has indicated that
the correct value for $\gamma$ should be near to unity rather than the value
$0.5$ used above in which case the ratio is considerably increased and there
may no longer be any indication of $\Delta I={1\over 2}$ 
rule violation.\cite{41}

\section{Conclusions}
 
The subject of hypernuclear weak decay has recently undergone somewhat of a
renaissance.  During the early 1960's there existed a healthy combination of
emulsion based experiments and theoretical work by Dalitz' Oxford group and
by Adams at Stanford.  Then for nearly two decades there was a dearth of 
activity in this area.  The situation changed during the 1980's, with the
fast counting experiments mounted at Brookhaven by the 
CMU-BNL-UNM-Houston-Texas-Vassar collaboration,
which renewed interest in this subject.  The work has generated a good deal
of theoretical activity, as described above, as well as a variety of new
and proposed experiments.

Above we described our own calculational program, which utilizes a meson
exchange picture of the $\Lambda N\rightarrow NN$ process in hypernuclei.
Because we are dealing with a $\Delta S=1$ weak transition, there exist a 
variety of mesons which must be considered, both strange and non-strange.  
The weak BB'M vertices were evaluated using the same SU(6)$_w$
and quark model techniques which have been utilized with some success in
the treatment of the somewhat similar problem of nuclear parity violation.
In comparing the predictions of our calculations and those of other groups
with the limited experimental data, it must be acknowledged that
the present situation is unsatisfactory.  Although
there is very rough qualitative agreement between theoretical expectations and
experimental measurements, it is not clear whether those discrepancies which do 
exist are due to experimental uncertainties, to theoretical insufficiencies,
or both.  On the theoretical side, what is needed are reliable calculations
on finite hypernuclei (preferably by more than one group) 
which clearly indicate the signals that 
should be sought in the data.  The issue associated with the validity
of the $\Delta I={1\over 2}$ rule must be clarified.  In addition there have
been recent speculations about the possible significance 
of two-nucleon stimulated 
decay\cite{42} (which could account for as much as 15\% of the decay amplitude
according to estimates) and of the importance of direct quark 
(non-meson-exchange) mechanisms,\cite{43}
which deserve further study in order to eliminate the vexing double counting
problems which arise when both are included.  Proper treatment of the former
involves a careful treatment of the many-body aspects of the nuclear medium
in the presence of the weak interaction and will require study by nuclear
theorists.  On the other hand, the sorting out of direct quark vs.
meson-exchange components of the weak amplitude is an extremely subtle problem 
requiring the efforts of both particle and nuclear theorists.  For example,
recent works by Maltman and Shmatikov\cite{44} and by Inoue, Takeuchi and
Ota\cite{45} calculate so-called ''direct" four-quark contributions to weak
hypernuclear decay by evaluating the (renormalization group corrected)
local Hamiltonian in the context of a simple quark model for the baryon
states.  The problems with such an approach are at least two.  Firstly, as
discussed earlier in this paper, this technique cannot account for the validity
of the $\Delta I={1\over 2}$ rule in other contexts---without {\it soft} gluonic
effects properly included one cannot expect this simple picture to generate a
realistic picture of hypernuclear decay.  The second problem is that the
''direct" mechanism omits the important contributions associated with meson
exchange, as calculated in our work.  In a quark picture these are associated
with diagrams such as that shown in Figure 9.  In order to avoid double
counting issue, such pieces must be carefully
subtracted from any ''direct" quark calculation before combining ''direct" 
and meson exchange components.  This is a highly non-trivial
assignment which has yet to be solved, and which we relegate to future
work.
\begin{figure}
\vspace{3.5in}
\caption{Quark picture of meson exhange contribution to nonmesonic hyperon
decay.}
\end{figure}

On the experimental side, we 
require 
an extensive and reliable data base developed in a variety of nuclei in order 
to confirm
or refute the predicted patterns.  The existence of new high intensity 
accelerator facilities
such as CEBAF and DA$\Phi$NE will be important in this regard.  In any case
it is clear that hypernuclear weak decay is a field in its infancy and that
any present theoretical optimism may well all too soon be tempered by experimental
reality.

\end{document}